\documentclass[%
 reprint,
superscriptaddress,
showpacs,preprintnumbers,
footinbib,
 amsmath,amssymb,
pra,
floatfix,
]{revtex4-1}

\usepackage[T1]{fontenc}
\usepackage[utf8]{inputenc}
\usepackage{amsmath}
\usepackage{amsthm}
\usepackage{amsfonts,dsfont}
\usepackage{graphicx}

\usepackage{bbm}
\usepackage{hyperref}
\usepackage{color}
\usepackage{comment}

\theoremstyle{theorem}
\newtheorem{proposition}{Proposition}

\newcommand{\bra}[1]{\langle #1 |} 
\newcommand{\ket}[1]{| #1 \rangle } 
\newcommand{\upd}{\mathrm{d}}
\newcommand{\tr}{\mathrm{tr}}
\newcommand{\ie}[0]{\textit{i.e.} }
\newcommand{\eg}[0]{\textit{e.g.} }

\definecolor{cbl}{rgb}{0,0,1}

\definecolor{crd}{rgb}{1,0,0}

\begin{document}
\title{Spikes in quantum trajectories}

\author{Antoine Tilloy}
\email{antoine.tilloy@ens.fr}
\affiliation{Laboratoire de Physique Th\'eorique,
CNRS and Ecole Normale Sup\'erieure de Paris, France
}
\author{Michel Bauer}
\email{michel.bauer@cea.fr}
\affiliation{
Institut de Physique Th\'eorique, CEA Saclay and CNRS, Gif-sur-Yvette, France
}
\author{Denis Bernard}
\email{denis.bernard@ens.fr}
\affiliation{Laboratoire de Physique Th\'eorique,
CNRS and Ecole Normale Sup\'erieure de Paris, France
}


\begin{abstract}
A quantum system subjected to a strong continuous monitoring undergoes quantum jumps. This very well known fact hides a neglected subtlety: sharp scale-invariant fluctuations invariably decorate the jump process even in the limit where the measurement rate is very large. This article is devoted to the  quantitative study of these remaining fluctuations, which we call \emph{spikes}, and to a discussion of their physical status. We start by introducing a classical model where the origin of these fluctuations is more intuitive and then jump to the quantum realm where their existence is less intuitive. We compute the exact distribution of the spikes for a continuously monitored qubit. We conclude by discussing their physical and operational relevance.
\end{abstract}
\maketitle

\section{Introduction}
The quantum jumps emerging from the continuous monitoring of a quantum system have been known since the begining of quantum mechanics \cite{bohr1913} and have been the subject of both theoretical \cite{wiseman2012,jordan2013,lmp} and experimental \cite{nagourney,bergquist,sauter,weber2014} investigation. It has been emphasized recently \cite{jumps}, though it was certainly expected, that they are a generic phenomenon in the sense that they necessarily appear anytime a quantum system is subjected to a strong continuous monitoring. However, fluctuations around the dominant jump process could already be descried in the early numerical work on the subject \cite{gisin1992,mabuchi1998}. They persist in the strong measurement limit and are not artifacts of experimental uncertainties, a fact which seems to have been largely overlooked in the literature. Even for strong (diffusive) continuous measurements, the Poissonian jumps of the density matrix always appear decorated with a residual noise. When the measurement rate becomes infinite, this decorating noise becomes punctual and has a vanishing impact on the finite dimensional probability distributions. However, as we will see, this does not mean that it can be discarded altogether.

The objective of this article is to prove that some sharp fluctuations, looking like aborted jumps and which we call \emph{spikes}, indeed persist even in the infinitely strong measurement limit and to show that they can be precisely quantified. We start by studying a classical toy model of iterated imperfect measurement where the phenomenon of spikes appears in a very clear and non puzzling way. We then go to the quantum realm where we show that a continuously and perfectly monitored qubit displays the same type of fluctuations. We compute the distribution of the spikes in these two cases and use the first to provide a physical intuition on the second.

\section{A classical toy model}

Let us start with a model for a classical iterated imperfect measurement. We consider a \emph{classical} particle hopping between two compartments "left" and "right" of a box. To make things simpler we start with a discrete time and assume that the particle has a probability $\lambda$ to change of compartment at every step. Our objective is to track the \emph{real} position $R_t$ of the particle ($R_t=1$ for "left" and $R_t=0$ for "right"). For that matter we take a photo of the box at each time step but we assume that the camera is bad and provides us only with very blurry pictures. 
\begin{figure}[b]
\includegraphics[width=\columnwidth,trim = 0cm 20cm 0cm 0cm, clip]{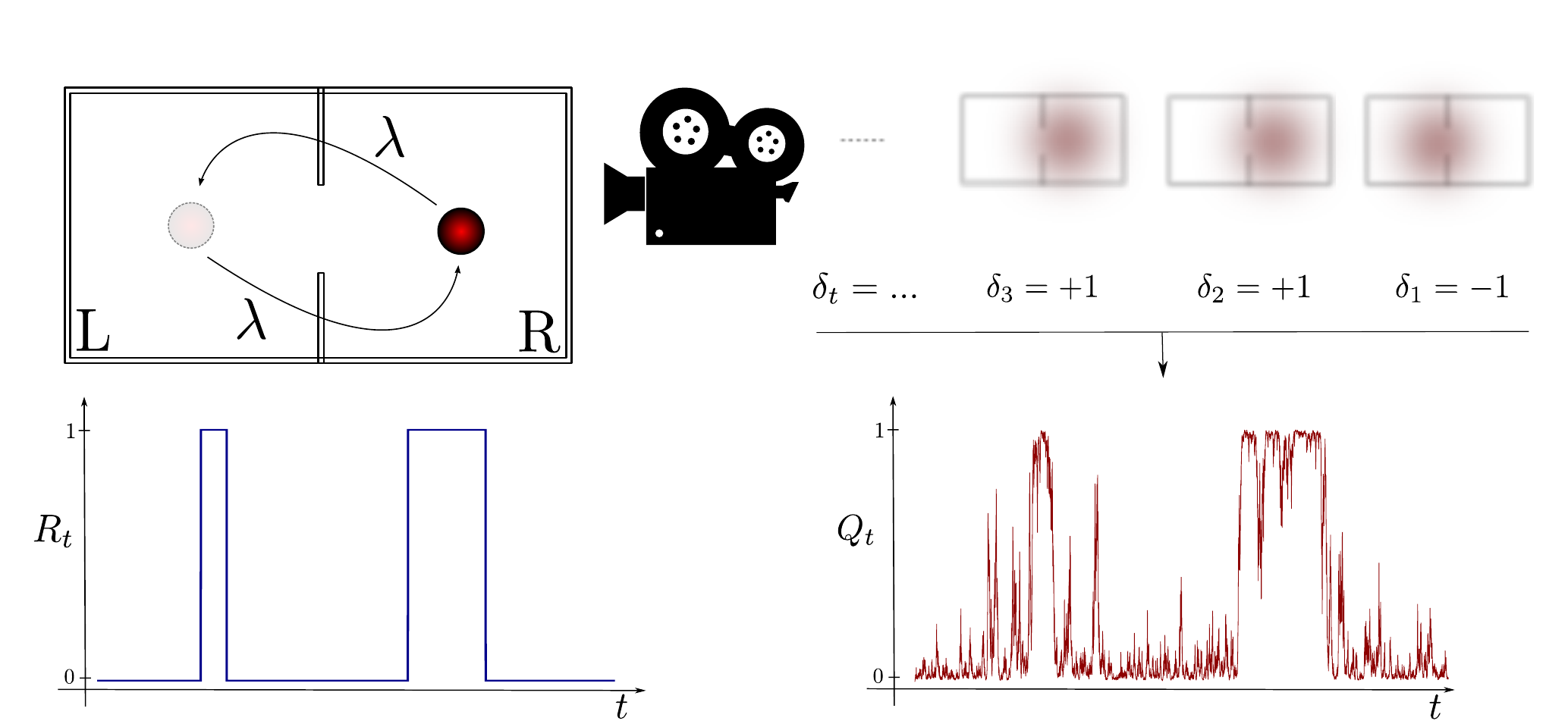}
\caption{Schematics of the classical model considered. The particle jumps are recorded in a variable $R_t$, an imperfect camera yields blurry photos from which a binary variable $\delta_i$ is extracted. An estimate $Q_t$ of $R_t$ is constructed from the $\delta_u$'s for $u<t$. The obtained $Q_t$ is expected to be close to $R_t$ when the camera takes many images per unit time.}
\label{fig:schematics}
\end{figure}
Every photo yields a binary answer $\delta_n=\pm 1$ which gives some information about the possible localization of the particle:
\begin{equation}
\begin{split}
&\mathds{P}(\delta_n=1 | \text{particle on the left})=\frac{1+\varepsilon}{2}\\
&\mathds{P}(\delta_n=1 | \text{particle on the right})=\frac{1-\varepsilon}{2}
\end{split}
\label{eq:def}
\end{equation}
where $\varepsilon \in\, ]0,1[$ represents the measurement precision. What we have constructed (see Fig. \ref{fig:schematics}) is actually one of the simplest possible instance of a Hidden Markov Model \cite{bechhoefer2015}. As often in such a situation, we are interested in knowing the best estimate $Q_n$ that the particle is on the left at the $n$-th iteration, i.e.:
\[
\begin{split}
Q_n &= \mathds{P}(\text{particle on the left at time }n | \text{pictures before } n)\\
&= \mathds{P}(R_n=1|\lbrace\delta_k\rbrace_{k\leq n})
\end{split}
\]
To compute $Q_{n+1}$, knowing $Q_n$ and the result from the last picture $\delta_{n+1}$, we simply need to:
\begin{itemize}
\item Incorporate the information from the last measurement result $\delta_{n+1}$ using \emph{Bayes rule}:
\begin{equation}
\begin{split}
Q_{n+1}&=\mathds{P}(\text{left at n+1}| \lbrace\delta_k\rbrace_{k\leq n}, \delta_{n+1})\\
&=\frac{\mathds{P}(\delta_{n+1}|R_{n+1}=1)\,\mathds{P}(R_{n+1}=1|\lbrace\delta_k\rbrace_{k\leq n})}{\mathds{P}(\delta_{n+1}|\lbrace\delta_k\rbrace_{k\leq n})}
\end{split}
\label{eq:bayes}
\end{equation}
\item Take into account the fact that we know that the particle tends to jump during the time interval separating two photos:
\begin{equation}
\begin{split}
&\mathds{P}(R_{n+1}=1|R_{n}=1)=1-\lambda\\
&\mathds{P}(R_{n+1}=1|R_{n}=0)=\lambda
\end{split}
\end{equation}
\end{itemize}
Using the law of total probability and the fact that $\mathds{P}(R_{n+1}|R_n)$ is independent from $\lbrace\delta_k\rbrace_{k\leq n}$ gives
\begin{equation}
\mathds{P}(R_{n+1}=1|\lbrace\delta_k\rbrace_{k\leq n})=(1-\lambda)Q_n + \lambda (1-Q_n).
\label{eq:evolution}
\end{equation}
Similarly the law of total probability can be used to compute the denominator of equation (\ref{eq:bayes}). Eventually, inserting formulae (\ref{eq:def}) and (\ref{eq:evolution}) in equation (\ref{eq:bayes}) one can construct $Q_{n+1}$ from $Q_n$ and $\delta_{n+1}$:
Elementary algebra and standard probability theory then give:
\begin{equation}
Q_{n+1}=\frac{(1+\varepsilon\delta_{n+1})\left[(1-\lambda)Q_n+\lambda(1-Q_n)\right]}{1+2\varepsilon\delta_{n+1}\left[(1-\lambda)Q_n+\lambda(1-Q_n)-1/2\right]}
\label{eq:final}
\end{equation}
\ie $Q_n$ can be iteratively constructed from $\lbrace\delta_k\rbrace_{k\leq n}$. To summarize we have a \emph{real} physical quantity $R_n$ that jumps between $0$ and $1$. The only information available at step $n$ is the collection of $\delta_k$ for $k\leq n$ from which we can construct, the best estimation $Q_n$ of $R_n$ at the step $n$. Our objective is to show the similarities and differences between the behavior of $R_n$, the physical quantity, and $Q_n$, which represents what we \emph{know} of $R_n$. We can loosely say that we have a lot of information on $R_n$ if $Q_n$ is close to $0$ or $1$.

For computational convenience, one can derive the continuous version of the previous discrete evolution in the limit of infinitely blurry pictures taken at an infinitely high frequency (so that information is extracted continuously at a finite rate). With the following rescaling and definitions:
\begin{equation}
\begin{array}{lll}
 t=n \delta t, & \varepsilon =\sqrt{\gamma} \sqrt{\delta t}/2, &\lambda=\tilde{\lambda}\delta t/2,
\end{array}
\end{equation}
one can write a stochastic differential equation\footnote{Incidentally, eq. (\ref{eq:sde}) can also be seen as describing the continuous energy monitoring of a qubit coupled to a thermal bath at infinite temperature (see \eg \cite{lmp}), in that case, $Q_t$ is the probability to find the qubit in the ground state). This means that the features of this classical toy model can also be displayed by genuinely quantum systems.} (in the It\^o form) for the evolution of $Q_t$ as seen from the observer's perspective:
\begin{equation}
\upd Q_t=\tilde{\lambda} \left(\frac{1}{2}- Q_t\right) \upd t + \sqrt{\gamma}\;Q_t (1-Q_t) \upd W_t,
\label{eq:sde}
\end{equation}
where $W_t$ is a Wiener process (i.e. $\frac{\upd W_t}{\upd t}$ is the Gaussian white noise). The details of the derivation can be found in appendix \ref{sec:continuum}. Let us emphasize the fact that this continuous limit is only needed to get closed form results and to give a more intuitive understanding of the phenomena at play. It is by no means required for the "spikes" to emerge. Even with a fixed value of $\varepsilon<1$ \ie even for $\varepsilon$ close to $1$, numerical simulations show that the plots look qualitatively the same for the large $\gamma$ limit we are going to consider next; the results we are going to show are not an artifact of the rescaling. 

Equation (\ref{eq:sde}) is now easy to interpret with the physical picture in mind. When $\gamma$ is small, \textit{i.e.} when the rate at which information is extracted is negligible compared to the jump rate, the evolution is dominated by the drift term which tends to drag the probability to $1/2$, that is to complete ignorance. When $\gamma$ is large, the noise term dominates and $Q_t$ tends to be attracted to the fixed points of the diffusion, $0$ and $1$, \textit{i.e.} perfect certainty. This intuition is largely confirmed by a direct numerical simulation of the evolution (see Fig. \ref{fig:comparison}). However, it misses an important aspect of the fluctuations. 

The numerical simulations of Fig. \ref{fig:comparison} show that our estimate $Q_t$ indeed undergoes jumps mirroring those of $R_t$ when $\gamma$ is large. What is more surprising is the fact that the sharp fluctuations around the plateaus do not disappear for large $\gamma$ as one would naively expect. They become sharper and sharper but their statistics do converge to a limit. These remaining net fluctuations, which become instantaneous when $\gamma \rightarrow \infty$, are what we call \emph{spikes}. Spikes can be characterized by the following proposition:

\begin{figure}
\includegraphics[width=0.95\columnwidth,trim = 0cm 14.7cm 0cm 0cm, clip]{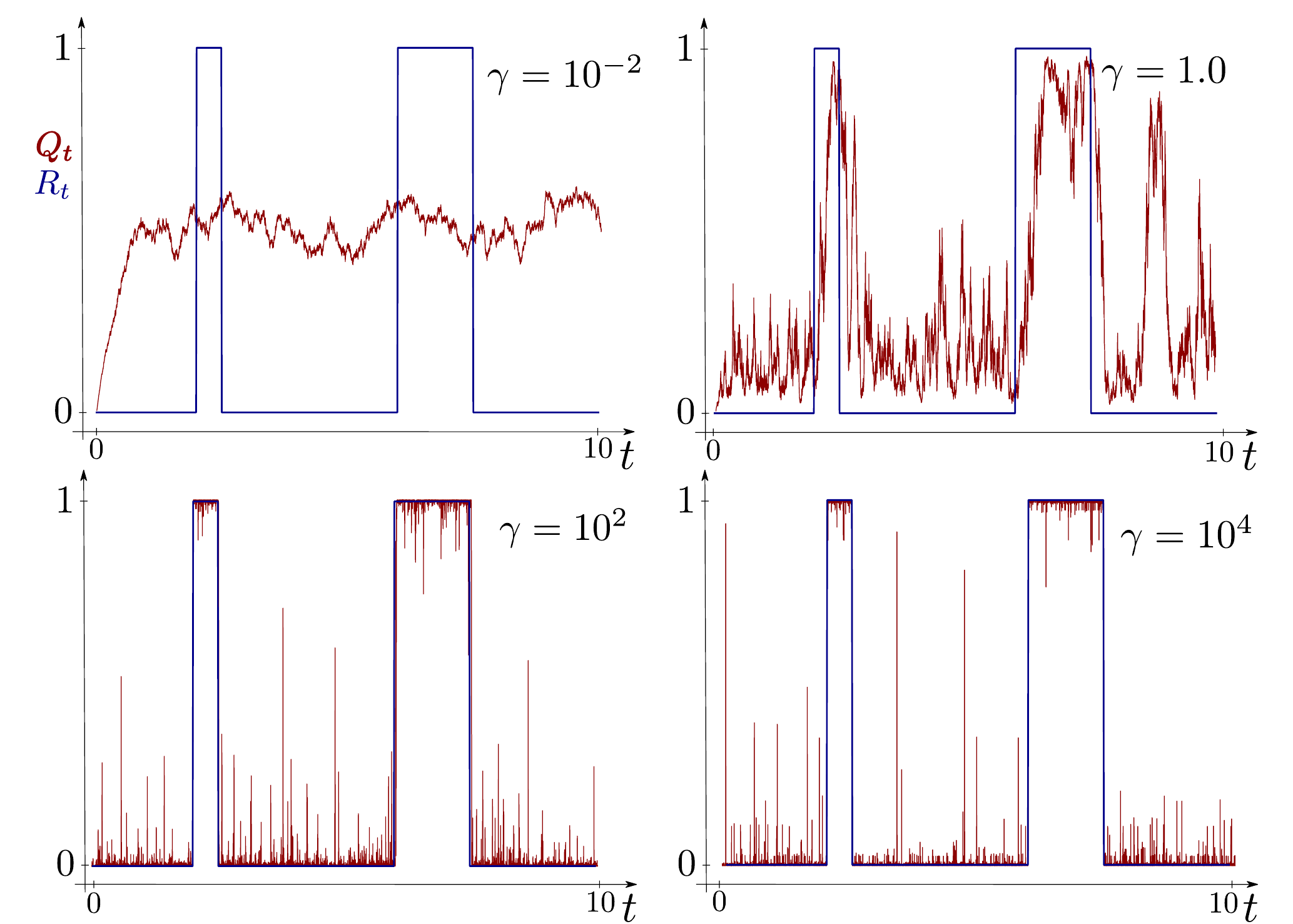}
\caption{Emergence of the spikes: $Q_t$ for various choices of $\gamma$. Top-left $\gamma=10^{-2}$, the information flow is weak and $Q_t$ fluctuates around $1/2$. As $\gamma$ increases, $Q_t$ gets closer to $R_t$ but sharp excursions persist. No qualitative difference can be seen with the naked eye between $\gamma=10^4$ and $\gamma \rightarrow +\infty$.}
\label{fig:comparison}
\end{figure}
\begin{proposition}
In a time interval [0,T] when $R_t=0$, the probability to have $N$ spikes with maxima in the domain $D\subset [0,T]\times [0,1[$ is a Poisson variable, i.e.
$\mathds{P}(N)= \frac{\mu^N}{N!}e^{-\mu}$ with intensity $\mu =\int_D d\nu$. The measure $d\nu$ is given by:
\begin{equation}
\upd\nu=   \frac{\tilde{\lambda}}{Q^2}\;\upd Q \upd t
\label{eq:nu}
\end{equation}
\end{proposition}
\begin{figure}
\centering
\includegraphics[width=0.6\columnwidth,trim = 0cm 0.15cm 0cm 0cm, clip]{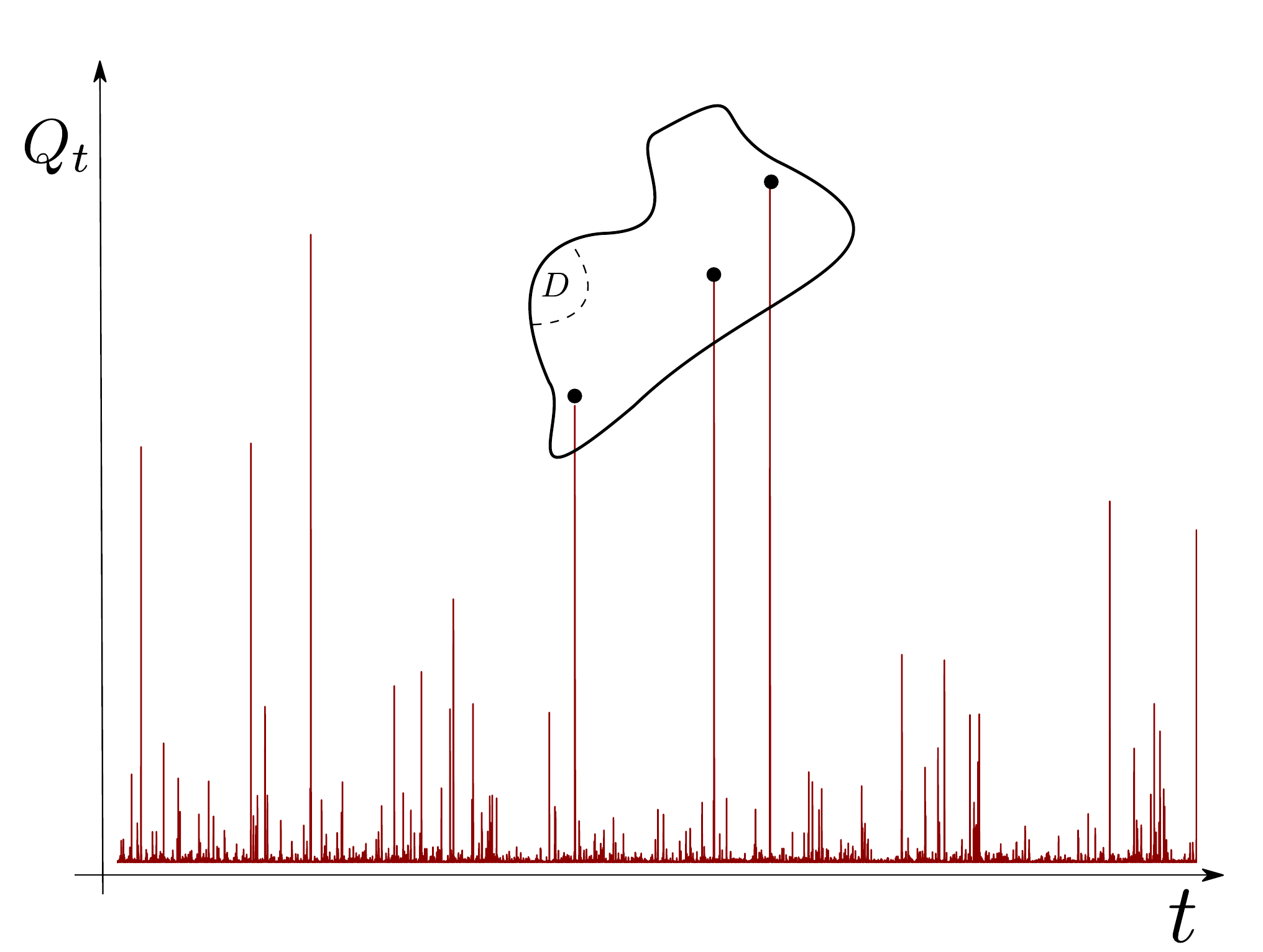}
\caption{Details of the spikes in an interval where $R_t=0$. The number of spikes (here 3) in the domain $D$ is quantified by the proposition.}
\label{fig:domain}
\end{figure}
The situation is completely symmetric when $R_t=1$, the spikes then start from the top of the graph, \textit{i.e.} $Q=1$ and go down; in this case $\upd\nu=   \frac{\tilde{\lambda}}{(1-Q)^2}\;\upd Q \upd t$.
The proof is provided in appendix \ref{sec:prooftoymodel} and extensively uses the fact that when $\gamma$ is large, $Q_t$ is \emph{almost} a continuous martingale, \emph{i.e.} a Brownian motion up to a time re-parametrization \cite{oksendal1992,feller2008}.

Let us remark a few straightforward consequences of the proposition. Spikes are scale invariant fluctuations. More precisely, Fig. \ref{fig:domain} would look exactly the same with the transformations $t\rightarrow A t$ and $Q\rightarrow AQ$. There are consequently infinitely many small spikes (but a finite number of spikes bigger than $Q_0$ for $Q_0 >0$).

This phenomenon of spikes could be thought to have no interesting practical consequences, after all as there is only a countable number of punctual spikes in the $\gamma\rightarrow +\infty$ limit, the simple jump process with the spikes removed is almost everywhere equal to the one dressed with the full fluctuations. However, if we look at quantities like arrival times, taking into account the spikes is fundamental. The probability that $Q_t$ reaches a region of relative uncertainty, say $0.4<Q_t<0.6$, in a given time interval has a dramatically higher value once the spikes are taken into account. As a more concrete example, consider two consecutive jumps of $R$ occuring at time $t_1$ and $t_2$ with $t_2-t_1 = \ln 2/\tilde{\lambda}$. Then we can see from the proposition that, in the interval $]t_1,t_2[$, we have $50\%$ chance to make at least one wrong prediction of $R_t$ using $Q_t$, \textit{i.e.} \[\mathds{P}\left(\exists t \in ]t_1,t_2[ , |Q_t-R_t|>\frac{1}{2}\right)=\frac{1}{2},\] 
even though strictly nothing physical happens inside this interval and the measurement rate is infinite! Had we naively taken the large $\gamma$ limit we would have imagined this probability to be 0.

\begin{figure}
\includegraphics[width=0.95\columnwidth,trim = 0cm 14.7cm 0cm 0cm, clip]{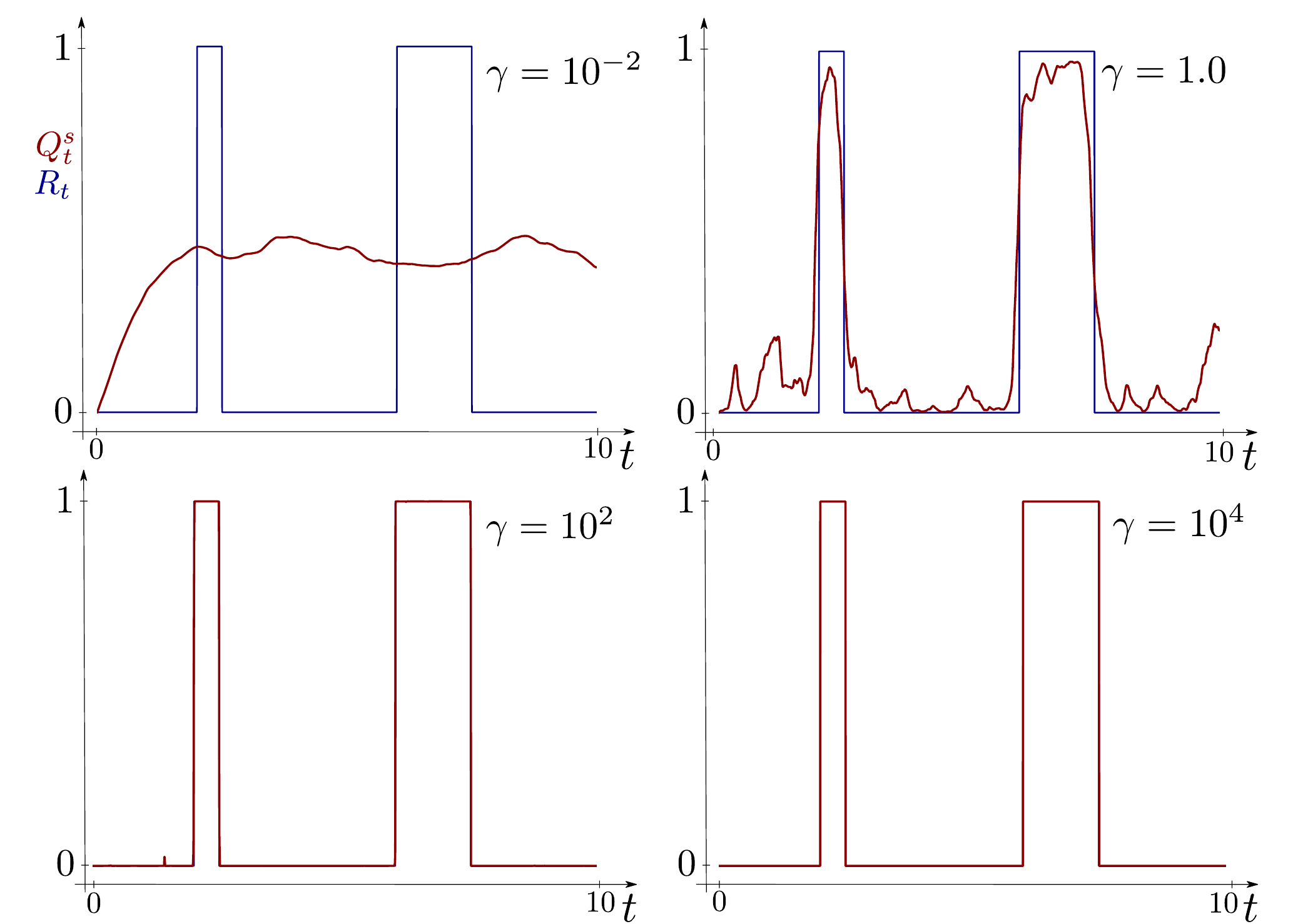}
\caption{The smoothed estimate $Q_t^s=\mathds{E}[R_t|\delta_u,u\in\mathds{R}]$ of $R_t$ shows no spikes.}
\label{fig:comparison_smoothed}
\end{figure}

However, even if the fine description of spikes is very useful in practice, it has no dramatic fundamental implications. No \emph{physical} quantity is intrinsically "spiky", only our information on the particle position behaves in such a peculiar way. Had we considered a forward-backward (or smoothed) estimate, \textit{i.e}. had we used all the photos to \emph{retrodict} the position of the particle at an earlier time, the estimate $Q_t^s=\mathds{E}[R_t|\delta_u,u\in\mathds{R}]$ would have been more regular  and notably spikeless (see Fig. \ref{fig:comparison_smoothed}). At most this means that the (filtered) Bayesian estimate of a quantity does not need to behave like the quantity itself even in the limit where the estimate is naively expected to be faithful. The emergence of a similar phenomenon in the quantum realm, where no underlying spikeless jump process can simply be invoked, should be more surprising.

\section{The quantum realm}

A good and widely studied (purely) quantum analog of our classical toy model is the continuous monitoring of a quantum system. For simplicity we will only treat the case of a 2-level quantum system but conjecture that the phenomenon we will see is ubiquitous in the sense that any continuously measured quantum system would display similar fluctuations.

\begin{figure}[b]
\includegraphics[width=0.99\columnwidth,trim = 0.8cm 20.5cm 2.5cm 0cm, clip]{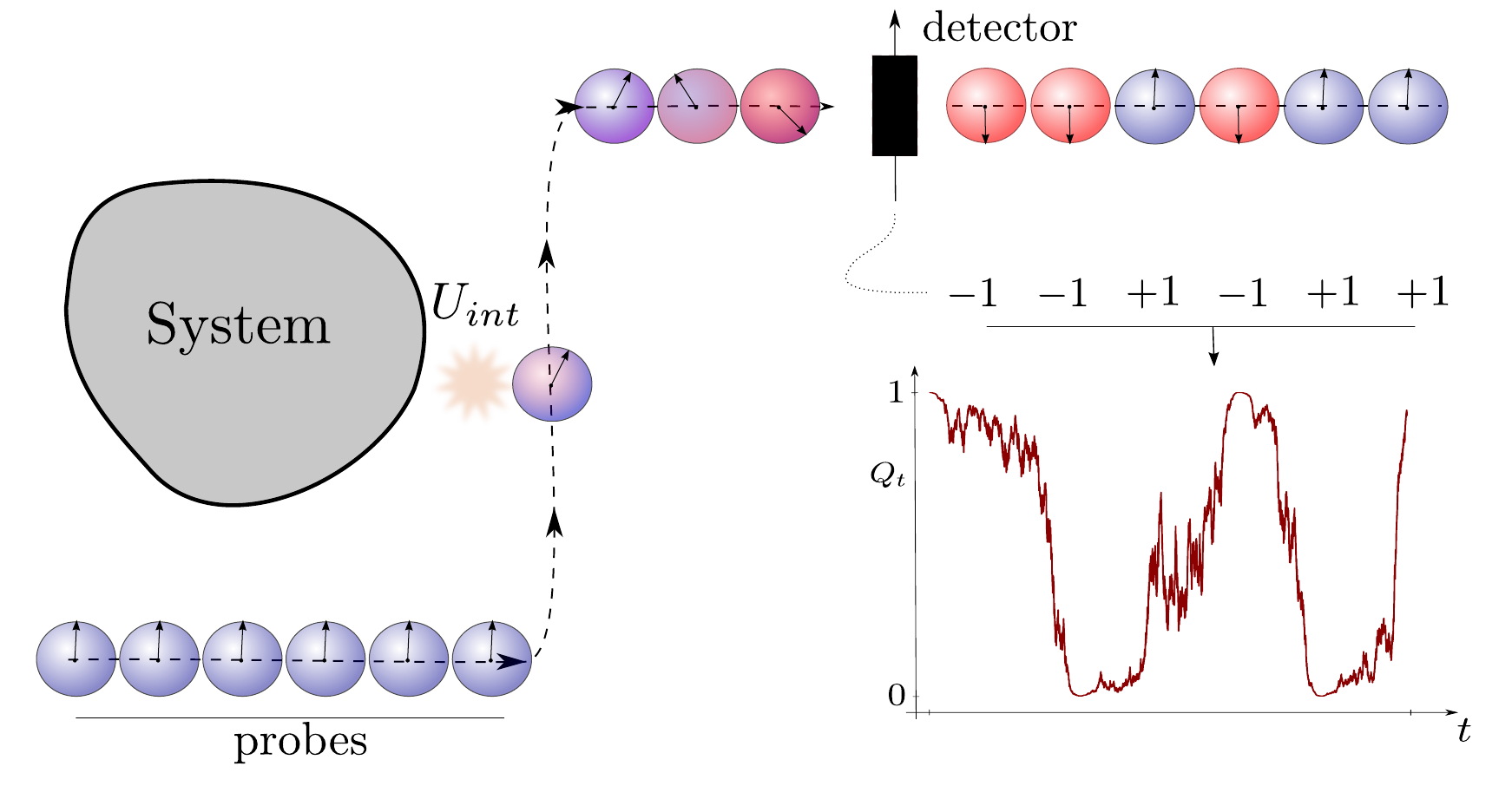}
\caption{Repeated interaction scheme. A collection of probes, typically two level systems, successively interact unitarily with the system before being projectively measured by a perfect detector. The probability $Q_t=\langle \psi |\rho_t|\psi\rangle$ for the system to be in some state $\ket{\psi}$ can be computed using the measurement results and the standard rules of quantum mechanics. Notice the analogy with the classical scheme of Fig. \ref{fig:schematics}. In this setting, the randomness in the evolution does not come from experimental imperfections but is purely of quantum origin.}
\label{fig:repeated_interaction}
\end{figure}

Let us consider the continuous monitoring of a qubit (e.g. a spin $1/2$ in a magnetic field). In analogy with the classical toy model, this can be seen as an iteration of weak measurements on a qubit \cite{attal2006,pellegrini2008,pellegrini2009,qbm} carried out via repeated interactions with probes. In such a setting a probe interacts for some time with the system before being measured by a perfect detector (see Fig. \ref{fig:repeated_interaction}). Because of the entanglement between the probe and the system, this measurement result gives some information on the system state. The probes consequently play exactly the same role as the blurry pictures of the classical case. Using the standard rules of quantum mechanics, it is straightforward to get the evolution of the system state $\rho$ as a function of time knowing the initial state of the probes, the unitary system-probe interaction $U_{int}$ and the measurement results. In what follows, we will go directly to the continuous limit of this scheme (\ie quickly repeated weak interactions) but again, the results qualitatively hold in the discrete case.

We suppose that we continuously monitor the qubit in a basis different\footnote{This is necessary to avoid a trivial evolution, for if the measurement operator and the Hamiltonian commute, the density matrix will simply be progressively projected on a measurement pointer state with no subsequent evolution.} from its energy basis (e.g. we continuously measure a spin in a direction orthogonal to that of the external magnetic field).
Using continuous quantum measurement theory \cite{barchielli1986,caves1987,diosi1988,barchielli1991,wiseman1996,belavkin1992,barchielli2009} one can get a stochastic master equation (SME) for the evolution of the density matrix $\rho_t$ of the qubit (with $\hbar=1$):
\begin{equation}
\begin{split}
\upd\rho_t=&-i\frac{\Omega}{2}[\sigma_y,\rho_t] \upd t - \frac{\gamma}{2}\left[\sigma_z\left[\sigma_z,\rho_t\right]\right]\upd t \\
&+ \sqrt{\gamma} \left(\sigma_z \rho_t + \rho_t\sigma_z-2\mathrm{tr}[\sigma_z\rho_t]\right) \upd W_t,
\end{split}
\label{eq:qbit}
\end{equation}
where we have considered for simplicity a measurement along $Oz$ and a magnetic field along $Oy$ and have assumed an efficiency of $1$ which guarantees that the noise if of purely quantum origin. The first term corresponds to the unitary part of the system evolution and the two other terms correspond respectively to the decoherence and localization (or collapse) induced by the measurement scheme. In this setting, the analog $X_t$ of the integrated information from the photos $\sum_{k<t}\delta_k$, often called the \emph{signal}, verifies $\upd X_t=2\sqrt{\gamma}\tr[\sigma_z \rho_t]\upd t+\upd W_t$. To avoid a complete Zeno freezing of the evolution, which will undoubtedly happen if we increase the measurement rate carelessly, we increase the magnetic field proportionally to the square root of the measurement rate, \textit{i.e.} we take
$\Omega=\sqrt{\gamma}\omega$ with $\omega$ constant. As before we are interested in the quantity $Q_t=\bra{+}_z\rho_t\ket{+}_z$ the probability that the qubit is found in the state $\ket{+}_z$ .

\begin{figure}
\includegraphics[width=0.95\columnwidth,trim = 0cm 14.5cm 0cm 0cm, clip]{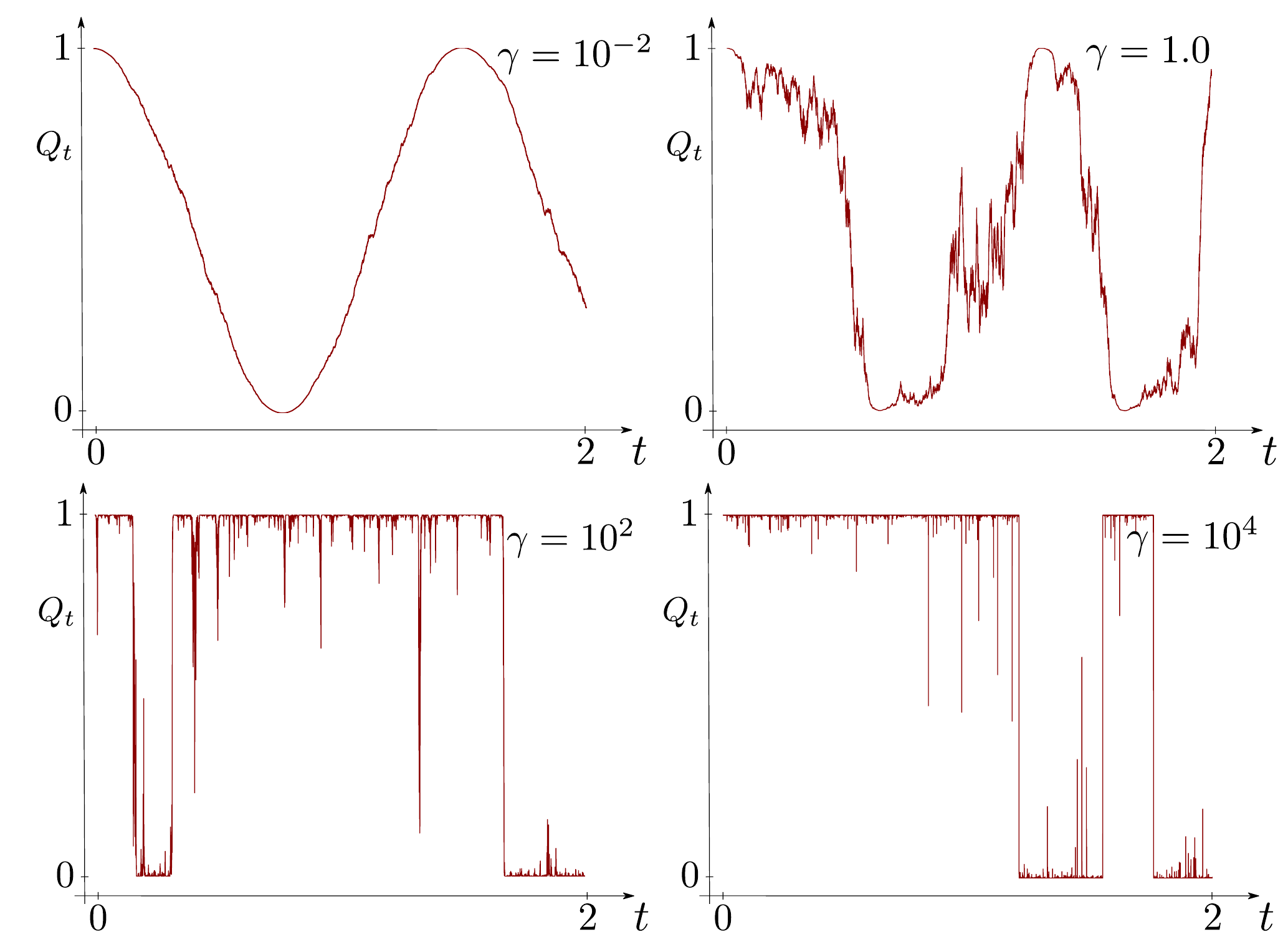}
\caption{Emergence of the spikes: $Q_t$ for various choices of $\gamma$. Again, no qualitative difference can be seen with the naked eye between $\gamma=10^4$ and $\gamma \rightarrow +\infty$. We take $\Omega = 2 +\sqrt{\gamma}$, \ie $\Omega$ \emph{not} exactly linear in $\gamma$ to keep a non trivial evolution (some Rabi oscillations) at $\gamma \simeq 0$. Notice that this time there is no underlying jump process $R_t$.}
\label{fig:comparison2}
\end{figure}

The direct numerical simulations of eq. (\ref{eq:qbit}) for various values of $\gamma$ are shown on Fig. \ref{fig:comparison2}. As before, we see the emergence of (quantum) \emph{spikes} around a jumpy trajectory which --at first sight-- seem to have the same characteristic than that of Fig. \ref{fig:comparison}. Actually, the analogy is not just qualitative and the spikes can be shown to be characterized by the same kind of Poisson process as before. This time the proposition reads:
\begin{proposition}
In a time interval [0,T] where $Q_t$ is close to 0, the probability to have $N$ spikes with maxima in the domain $D\subset [0,T]\times [0,1[$ is a Poisson variable, i.e.
$\mathds{P}(N)= \frac{\mu^N}{N!}e^{-\mu}$ with intensity $\mu =\int_D d\nu$. The measure $d\nu$ is given by:
\begin{equation}
\upd\nu=   \frac{\omega^2}{Q^2}\;\upd Q \upd t
\label{eq:nubis}
\end{equation}
\end{proposition}
The notion of being close to $0$ (or $1$) becomes well defined when $\gamma\rightarrow+\infty$ where $Q_t=0$ or $Q_t=1$ almost surely with a finite jump rate between the two cases. 
As in the case of the toy model, the situation is perfectly symmetric when $Q_t$ is close to $1$. The proof uses essentially the same method as before and is provided in appendix \ref{sec:proofqubit}.

With the help of numerical simulations, we conjecture that the phenomenon of spikes is ubiquitous in tightly monitored quantum systems. However, it is less clear to see along which direction of the density operator space the spikes should occur in a more general setting. This is a question which should be studied further and a fully rigorous proof of the generality of the spikes, completed by their precise characterization, would be gripping.

For the classical toy model, we have shown that using smoothing instead of filtering would remove the spikes. In the absence of a well defined hidden state in the quantum case, such a procedure could at least provide us with a spikeless quantity. Different quantum analogues of smoothing have been proposed in the literature \cite{Tsang2009,Gammelmark2013,Guevara2015} and already applied in experiments \cite{rybarczyk2014,campagne2014}. In \cite{Gammelmark2013}, the authors define the \emph{past quantum state} which is a quantum equivalent of smoothing rooted in the weak value formalism. The quantities computed with the \emph{past quantum state}, like the past density matrix, are more regular than $\rho_t$ and it can be checked numerically that they are spikeless. However because it is based on the weak value, this formalism suffers from the same subtleties of interpretation in the general case \cite{aharonov1988}. In an other interesting proposal \cite{Guevara2015}, the authors define a procedure called \emph{quantum state smoothing} which allows to recover some of the information lost in an unmonitored bath via the use of future measurement results. This procedure keeps the standard interpretation of the quantum state. However, it cannot be used in our specific situation as it offers no gain for a measurement with efficiency 1 and no unmonitored decoherence channel: in this case there is, strictly speaking, no information to recover. The different proposals for a quantum equivalent of smoothing have various advantages and drawbacks and their ability to trim the spikes can give a criterion --among many others-- to compare their benefits.

\section{Discussion}
We have shown that the phenomenon of \emph{spikes} could be understood as a pure artifact of Bayesian filtering in a classical toy model where nothing physical (in that case the particle classical position $R_t$) is spiky. In the quantum case, such a straightforward answer cannot be provided in the absence of a hidden variable. Moreover, in the case we have considered, the quantum state stays \emph{pure} \footnote{This can be seen by computing $\upd(\det \rho_t)$ using It\^o formula and noticing that $\det \rho_t$ is a supermartingale converging exponentially fast to $0$ so that the evolution purifies the state. A general proof, valid in arbitrary dimension, is provided for example in \cite{maassen2006}} during the whole evolution making the spikes difficult to dismiss as a spurious consequence of ignorance. Eventually, the essence of the problem lies in the ``reality'' one is ready to grant the quantum state. Without going into these matters of interpretation, we believe the analogy with the classical case at least shows the quantum state behaves in an unintuitive way for a physical quantity. Nevertheless, spikes uncontroversially exist in the information we have of the system. Even if they have a vanishing support in the strong measurement limit, we have seen that they have an irreducible impact on quantities like arrival times. This means that spikes should absolutely be taken into account when designing feedback control schemes (where smoothing procedures cannot be applied). For example, a real-time control protocol requiring to do an operation every time the density matrix goes a bit too far from a predefined measurement eigenstate (as \eg in \cite{epl}) will be triggered very often --perhaps in some cases too often-- by spikes, something one would not have guessed taking naively the strong measurement limit.

An important problem is now the observation and quantification of spikes in experiments. Even if the observation of quantum trajectories is now done routinely by some skilled experimentalists \cite{gleyzes2007,murch2013,weber2014,de2014}, it is remarkable that spikes have not been conclusively pined down yet \footnote{Perhaps because spikes have often been mistaken for experimental errors in need for a fix and filtered out}. The simplest and already feasible way to observe and quantify spikes would be to look first at the discrete case, \ie at experiments of repeated interactions \eg as \cite{guerlin2007,hume2007}. Provided no ad-hoc averaging procedure is applied, one should be able to reconstruct the spikes. The difficulty is that the spikes show up no matter what the origin of the noise is and post processing procedures aimed at taming classical noise may very well suppress the quantum part as well. Finally, we have considered only simple two dimensional problems in this work. We have observed spikes numerically in any dimension but knowing more about their distribution in the general case would be illuminating.

\begin{acknowledgments}
This work was supported in part by the ANR contracts ANR-2010-BLANC-0414 and ANR-14-CE25-0003-01. A.T. would like to thank Irénée Frérot who suggested to study the classical toy model.
\end{acknowledgments}

\appendix
\section{Continuous limit of the toy model}\label{sec:continuum}
In this appendix, we provide a physicist proof of equation \eqref{eq:sde} of the main text, which is the limit of equation \eqref{eq:final} when $\lambda$ and $\varepsilon$ are small.
As we know that $\lambda$ will scale as $\upd t$ and $\varepsilon$ as $\sqrt{\upd t}$, we expand equation \eqref{eq:final} up to order $\upd t$ by keeping only terms of order $\varepsilon$, $\varepsilon^2$ and $\lambda$. This gives:
\begin{equation}\label{eq:discrete}
\begin{split}
Q_{n+1}-Q_n\simeq& \lambda (1-Q_n) - \lambda Q_n + 2\varepsilon Q_n (1-Q_n) \delta_{n+1} \\
&- 2 \varepsilon^2 Q_n (1-Q_n) (2Q_n-1)
\end{split}
\end{equation}
We now divide time $t$ in $m$ intervals $\upd t$ each of them divided in $l$ even smaller intervals $\delta t$, \ie $t=n \delta t=ml\delta t=m \upd t$. In what follow, both $l$ and $m$ will be large. We introduce $\upd Q_t=Q_{m(l+1)}-Q_{ml}$ and $\upd X_t= X_{(m+1)l}-X_{ml}=\sum_{k=ml}^{(m+1)l} \delta_{k}\sqrt{\delta t}$ and recall that
\begin{equation}
\begin{array}{lll}
\varepsilon =\sqrt{\gamma} \sqrt{\delta t}/2, &\lambda=\tilde{\lambda}\delta t/2,
\end{array}
\end{equation}
summing $l$ times equation (\ref{eq:discrete}) gives:
\begin{equation}
\upd Q_t\simeq\tilde{\lambda} (\frac{1}{2}-Q_t) \upd t +\sqrt{\gamma} Q_t (1-Q_t)\left(\upd X_t - \frac{1}{2}\sqrt{\gamma}(2Q_t-1) \upd t\right)
\label{eq:cont}
\end{equation}
When $l$ is large with $\upd t$ kept small, $\upd X_t$ is an (infinitesimal) Gaussian random variable as a sum of independent increments with (almost) the same law and equation (\ref{eq:cont}) becomes a stochastic differential equation. We only need to compute the mean and variance of the r.h.s of equation (\ref{eq:cont}) to rewrite it in the It\^o form.
\begin{equation}
\begin{split}
\mathds{E}(\upd X_t|\lbrace\delta_u\rbrace_{u\leq t})&\simeq\frac{\sqrt{\gamma}}{2}(2Q_t -1) l \delta t =\frac{\sqrt{\gamma}}{2}(2Q_t -1) \upd t\\
\mathds{E}(\upd X_t^2|\lbrace\delta_u\rbrace_{u\leq t})&\simeq l\delta t= \upd t
\end{split}
\end{equation}
Which means that $\upd X_t= \frac{\sqrt{\gamma}}{2}(2Q_t -1) \upd t + \upd W_t$ where $W_t$ is a Wiener process. Eventually we get:
\begin{equation}
\upd Q_t=\tilde{\lambda} \left(\frac{1}{2}- Q_t\right) \upd t + \sqrt{\gamma}\;Q_t (1-Q_t) \upd W_t,
\end{equation}
Which is what we have claimed.

\section{Proof of the emergence of the spikes}
\label{sec:proof}
In this appendix we provide an intuitive proof of the emergence of the spikes for the classical toy model and for the continuously monitored qubit.

\subsection{Case of the toy model}
\label{sec:prooftoymodel}
The basic ingredients are the fact that when $\gamma$ is large, $Q_t$ is \emph{almost} a martingale and the fact that stopped martingale, \textit{i.e.} a martingale conditioned on stopping at a predefined event,  is still a martingale. We will focus on the spikes starting from $Q=0$ (the situation is the same for $Q=1$). We will compute the statistics of the maximum reached by $Q_t$, starting from $q$ close to $0$ at t=0, before it reaches $q_\delta \ll q$. In what follows we will consequently have the hierarchy:
\begin{equation}
1> q \gg q_\delta \gg \gamma^{-1} \tilde\lambda
\end{equation}
\begin{figure}
\includegraphics[width=0.75\columnwidth,trim = 0cm 21.5cm 7.5cm 0cm, clip]{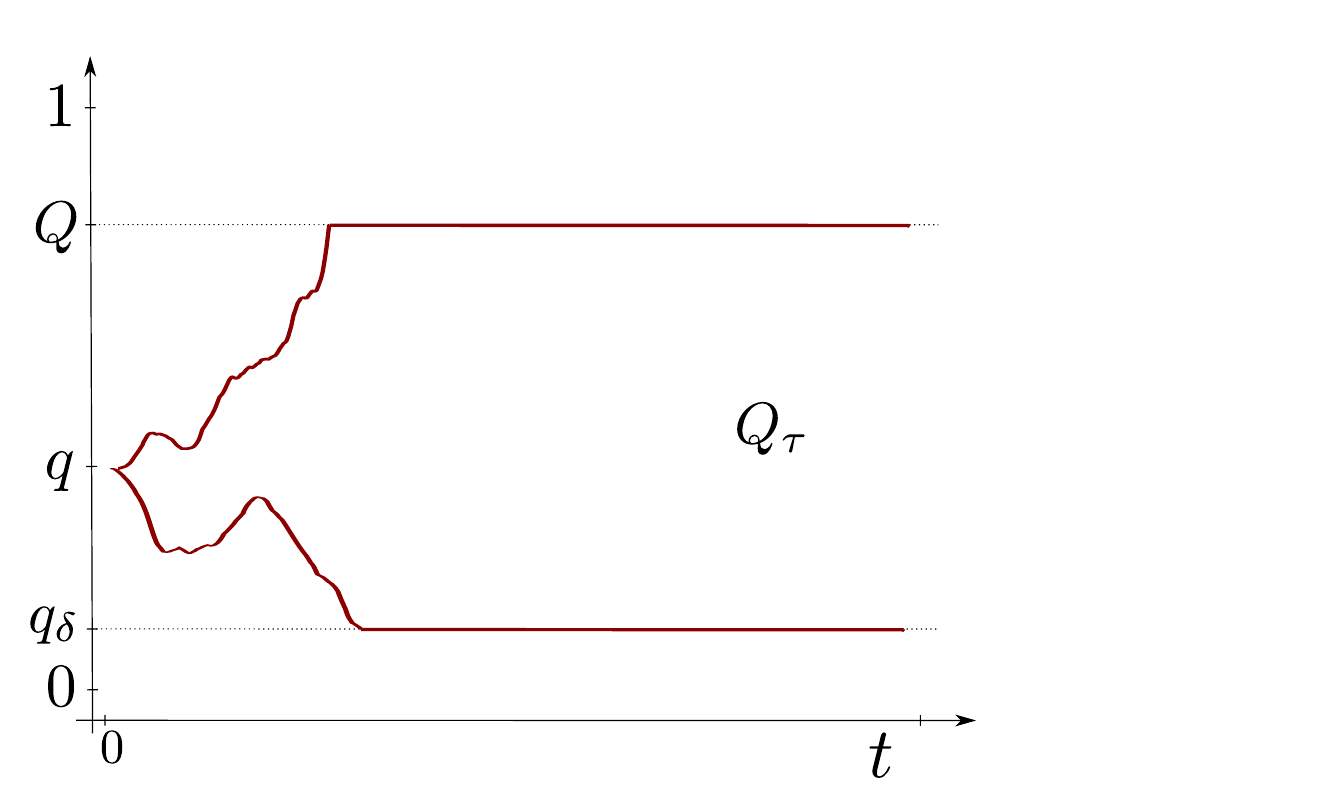}
\caption{Two trajectories of $Q_\tau$ starting from $q$.}
\label{fig:qtau}
\end{figure}
We consider the stopping time $\tau$ which is a random variable such that $\tau=t$ as long as $Q_t$ does not reach either $q_\delta$ or $Q$ and $\tau=t_{\text{reach}}$ after $Q_t$ reached $q_\delta$ or $Q$ at time $t=t_{\text{reach}}$. Because $1 \gg q_\delta \gg \gamma^{-1}\tilde \lambda$, we see from equation \eqref{eq:sde} that $Q_t$ is \emph{almost} a martingale, \emph{i.e.} the drift term is much smaller than the noise term. Probability theory now tells us that $Q_\tau$ (see Fig. \ref{fig:qtau}) is also a martingale which gives:
\begin{equation}
\mathds{E}(Q_\tau | Q_0=q)=q
\end{equation}
But the direct computation of this expected value for $t$ large gives:
\begin{equation}
\begin{split}
\mathds{E}(Q_\tau | Q_0=q)=&\mathds{P}(Q_\tau=Q| Q_0=q)Q \\
+& \mathds{P}(Q_\tau = q_\delta|Q_0=q) q_\delta  
\end{split}
\end{equation}
which gives: $\mathds{P}(Q_\tau=Q|Q_0=q) \simeq \frac{q}{Q}$. Something which we can rewrite:
\[
\mathds{P}\left(\underset{t, \forall u<t,\, Q_u > q_\delta}{\text{Max}}\left(Q_t\right) >Q|Q_0=q\right)=\frac{q}{Q}
\]
Or eventually in differential form:
\begin{equation} 
\upd\mathds{P}\left(\text{Max}\left(Q_t\right)=Q|Q_0=q\right)=q\left(\frac{\upd Q}{Q^2} + \delta(1-Q) \upd Q\right)
\label{eq:differentialform}
\end{equation}
This gives us the $1/Q^2$ dependency in equation (8). Notice also the Dirac mass, corresponding to complete jumps, that comes from the fact that $\mathds{P}\left(\underset{t, \forall u<t,\, Q_u > q_\delta}{\text{Max}}\left(Q_t\right) \geq 1|Q_0=q\right)=q \neq 0$. This equation is an equation for a conditioned probability and thus only gives us a ratio of probabilities, we still need to find the normalisation. This is naively given by the rate at which $Q_t$ reaches again $q$ after reaching $q_\delta$, \emph{i.e.} to how often $Q_t$ gets to "try" to do a large excursion. 

We first have to notice that an excursion, \ie starting from $q$, reaching a maximum $Q$ and going below $q_\delta$, takes an infinitely small amount of time in the large $\gamma$ limit. This simply comes from the fact that the dominant martingale term in equation \eqref{eq:sde} is independent from $\gamma$ once we do the rescaling $u=\gamma t$: $\upd X_u=Q_u(1-Q_u)\upd W_u$. This means that the typical time scale of an excursion is proportional to $\gamma^{-1}$, hence that excursions are instantaneous in the large $\gamma$ limit (which is why we now call them \emph{spikes}). What equation (\ref{eq:differentialform}) actually does in this context is giving us the probability that a spike higher than $q$ has a total height $Q$. 
\begin{figure}
\includegraphics[width=\columnwidth,trim = 0cm 21.5cm 0cm 0cm, clip]{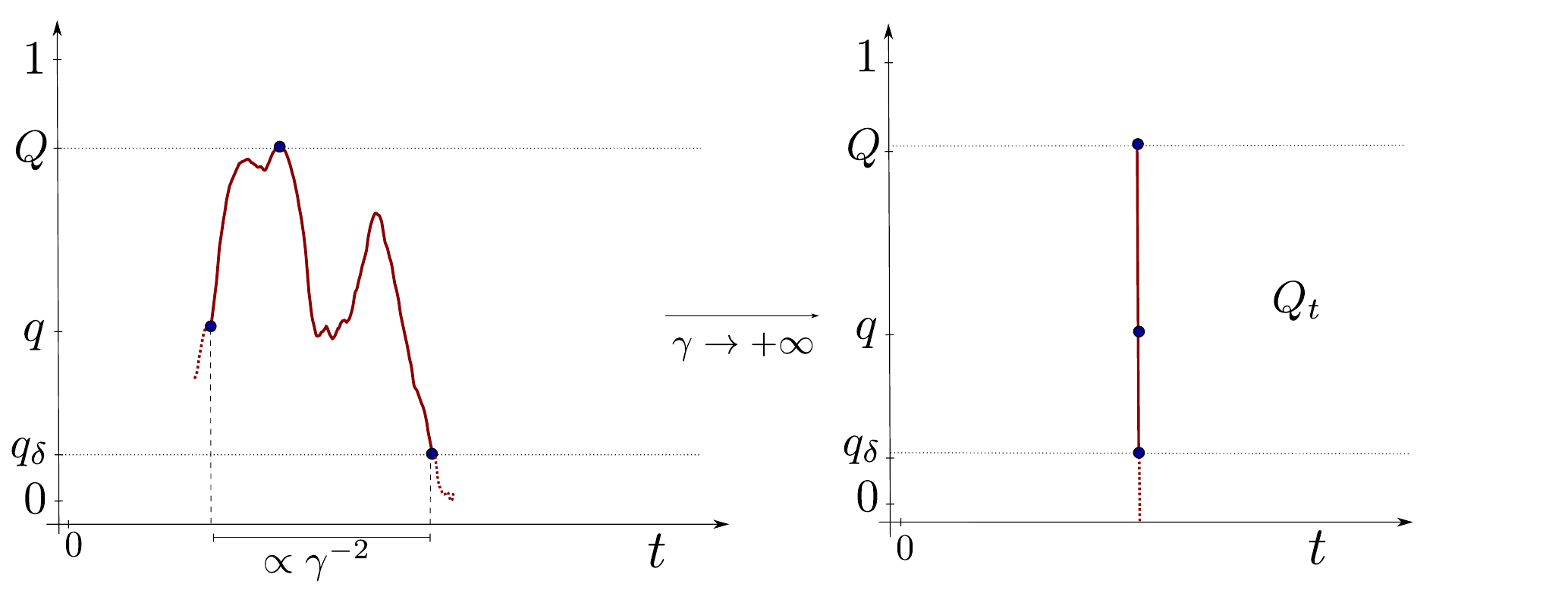}
\caption{On the left, a trajectory starting from $q$, going up to $Q$ and eventually going down below $q_\delta$. In the large $\gamma$ limit, this whole trajectory is followed almost instantaneously and looks like a single vertical spike (shown on the right).}
\label{fig:qspike}
\end{figure}
In an interval $\upd t$ we know just one thing, we know that the probability to reach $Q\simeq 1$, \ie the probability to do a complete jump, is $\tilde\lambda \upd t$. The probability that a spike higher than $q$ reaches $Q=1$ is simply $q$. This means that the probability to have a spike higher than $q$ in an interval $\upd t$ is $q^{-1}\tilde\lambda \upd t$, which provides the normalisation we wanted\footnote{We use the fact that as the probability to reach $q$ is infinitesimal, the probability to reach it twice or more is negligible}. Eventually, if we specify that we are on an interval without jumps to remove the Dirac mass, we get:
\begin{equation} 
\upd\mathds{P}\left(\text{Max}\left(Q_t\right)=Q\right)=\tilde\lambda \upd t\frac{\upd Q}{Q^2} 
\label{eq:differentialformnew}
\end{equation}
That is, during an infinitesimal time interval $\upd t$, the probability that there is a spike of maximum $Q$ up to $\upd Q$ is $\tilde\lambda \upd t \upd Q/{Q^2}$. As the spikes are independent from one another, this gives a Poissonian probability $\mathds{P}(N)$ to have $N$ spikes in any domain $D$:
\begin{equation}
\mathds{P}(N)= \frac{\mu^N}{N!}e^{-\mu},
\end{equation}
with $\mu =\int_D \frac{\tilde\lambda}{Q^2}\upd t \upd Q$, which is what we had claimed.

\subsection{Case of the qubit}
\label{sec:proofqubit}
The case of the qubit can be solved in the same way. Indeed, expanding equation (9) gives the following equation for $Q_t=\bra{+}_z\rho_t\ket{+}_z$ and $Y_t=\sqrt{\gamma}\bra{+}_z\rho_t\ket{-}_z$:
\begin{equation}
\begin{split}
\upd Q_t &= -\omega Y_t \upd t + 4 \sqrt{\gamma} Q_t (1-Q_t) \upd W_t\\
\upd Y_t &= \gamma \left[\omega (2Q_t-1)-2Y_t\right] \upd t +\sqrt{\gamma} Y_t(1-2Q_t)\upd W_t 
\end{split}
\label{eq:expandedqubit}
\end{equation}
We would like to use the same reasoning as before, \ie say that $Q_t$ is almost a martingale.
In order to do this, we have to show first that $Y_t$ is of order $0$ in $\gamma$. This is indeed the case: when $\gamma$ is large, $Q_t$ is typically close to $0$ or $1$. As before we will consider the case where $Q_t$ is close to $0$. In that case, and if we rescale time by taking $u=\gamma t$ we have:
\begin{equation}
\upd Y_u \simeq \left[-\omega -2Y_u\right] \upd u +Y_u \upd W_u 
\end{equation}
This means that the average of $Y$, $\bar Y_u$, verifies, the ordinary differential equation $\partial_u \bar Y_u = \left[-\omega -2\bar Y_u\right]$. When $\gamma$ is large, $u$ flows infinitely fast so that we reach the stationary value of $\bar Y$ instantaneously and $\bar Y_u \simeq \frac{-1}{2\omega}$. This means that the term $\omega Y_t $ is typically of order $0$ in $\gamma$ and thus negligible compared to the noise term in equation (\ref{eq:expandedqubit}).
This shows that, as before, $Q_t$ is almost a martingale in the large $\gamma$ limit (and is the same martingale as before) which is the only thing that was used in the first part of the previous proof.
What differs in this setting is the normalisation argument of the second part. It has been proved in \cite{jumps}, and it can be guessed by dimensional analysis, that the rate of (complete) jumps in the large $\gamma$ limit is $\omega^2$ in the case we now consider. This means that in the previous normalisation argument $\tilde \lambda \upd t$ simply needs to be replaced by $\omega^2 \upd t$ which provides the formula (10) we had put forward.

\bibliography{main}

\end{document}